\documentclass[twocolumn,epj]{svjour}  
\smartqed  

\usepackage{longtable}
\usepackage{grffile} 
\usepackage{graphics}
\usepackage{graphicx} 
\usepackage{amsmath}    
\usepackage{hyperref}   
\usepackage{subfigure}  
\usepackage{epsfig}
\usepackage{epstopdf}
\usepackage{color}
\usepackage{url}

\usepackage{float}
\usepackage{lineno}

\usepackage{xspace}
\usepackage{stfloats}

\newcommand{\vtwo}			{\ensuremath{v_{2}}\xspace}
\newcommand{\GeVc}          {Ge\kern-.1emV/$c$\xspace}
\newcommand{\TeV}           {Te\kern-.1emV\xspace}
\newcommand{\GeV}           {Ge\kern-.1emV\xspace}
\newcommand{\pt}           {\ensuremath{p_{\rm T}}\xspace}
\newcommand{\deta}           {\ensuremath{\Delta\eta}\xspace}
\newcommand{\ctwo}           {\ensuremath{c_2\{2\}}\xspace}
\newcommand{\cfour}           {\ensuremath{c_2\{4\}}\xspace}

\begin{document}

\title{Disentangling the development of collective flow in high energy proton proton collisions with a multiphase transport model}

\author{Liang Zheng\inst{1,5}  \thanks{{e-mail:} zhengliang@cug.edu.cn}
	\and Lian Liu\inst{1} 
	\and Zi-Wei Lin\inst{2} \thanks{{e-mail:} linz@ecu.edu}
	\and Qi-Ye Shou\inst{3} \thanks{{e-mail:} shouqiye@fudan.edu.cn}
	\and Zhong-Bao Yin\inst{4} \thanks{{e-mail:} zbyin@mail.ccnu.edu.cn}
}

\institute{School of Mathematics and Physics, China University of Geosciences (Wuhan), 
	Lumo Road 388, Wuhan 430074, China 
	\and
Department of Physics, East Carolina University,
Greenville, North Carolina 27858, USA
	\and
	Key Laboratory of Nuclear Physics and Ion-beam Application (MOE),
	Institute of Modern Physics, Fudan University, Shanghai 200433, China
	\and
Key Laboratory of Quark and Lepton Physics (MOE) and Institute
of Particle Physics, Central China Normal University, Wuhan 430079, China
	\and
Shanghai Research Center for Theoretical Nuclear Physics, NSFC and Fudan University, Shanghai 200438, China
}

\date{\\}

\abstract{ In this work, we investigate the collective flow development in high energy proton proton (pp) collisions with a multiphase transport model (AMPT) based on PYTHIA8 initial conditions with a sub-nucleon structure. It is found that the PYTHIA8 based AMPT model can reasonably describe both the charged hadron productions and elliptic flow experimental data measured in pp collisions at $\sqrt{s}=13$ TeV. By turning on the parton and hadron rescatterings in AMPT separately, we find that the observed collective flow in pp collisions is largely developed during the parton evolution, while no significant flow effect can be generated with the pure hadronic rescatterings. It is also shown that the parton escape mechanism is important for describing both the magnitude of the two-particle cumulant and the sign of the four-particle cumulants. We emphasize that the strong mass ordering of the elliptic flow results from the coalescence process in the transport model and can thus be regarded as unique evidence related to the creation of deconfined parton matter in high energy pp collisions.
}

\authorrunning{Liang Zheng et. al.}
\titlerunning{Disentangling the collective flow development in high energy proton proton collisions}
\maketitle

\section{Introduction}
\label{intro}

The exploration of hot and dense nuclear matter called quark gluon plasma (QGP) has been a long term quest for the heavy ion collision experiments performed at the Relativistic Heavy Ion Collider (RHIC) and the Large Hadron Collider (LHC)~\cite{Harris:1996zx,Elfner:2022iae,Busza:2018rrf}.
The elliptic flow (\vtwo), defined as the second-order Fourier component of the particle azimuthal distribution, is one of the most important experimental observables to uncover the properties of the QGP medium~\cite{Ollitrault:1992bk,Kolb:2003dz,Voloshin:2008dg}. The sizable \vtwo discovered in semi-central high energy heavy ion collisions is expected to be converted from the initial spatial anisotropy of the colliding system via final state interactions during the hydrodynamic expansion of the QGP droplet. Studying the collective flow in terms of its multi-particle correlations, transverse momentum (\pt) and hadron species dependence have been regarded as the unique opportunities to expose the evolution of the QGP matter produced in nucleus nucleus (AA) collisions~\cite{Shen:2011eg,Song:2017wtw,Heinz:2013th}. The experimentally observed signals like the sign change of four-particle cumulant (\cfour) and the mass ordering of flow harmonics are believed to be deeply rooted in the convoluted radial and elliptic hydrodynamic expansions of the QGP medium~\cite{Adamczyk:2013gw,Abelev:2014mda,Abelev:2014pua}.

In the past decade, collective flow like signals have also been surprisingly observed in numerous experimental data for small systems created in the high multiplicity proton nucleus (pA) collisions, proton proton (pp) collisions and even photon nucleus collisions which were not expected to flow~\cite{Khachatryan:2010gv,Abelev:2012ola,Khachatryan:2016txc,Adare:2014keg,Khachatryan:2015waa,PHENIX:2018lia,Acharya:2019vdf,ATLAS:2021jhn}. On aside of the continuing experimental efforts to further examine the resemblance between the small systems and the conventional large systems made by nucleus nucleus (AA) collisions, multiple theoretical frameworks have been constructed to understand the origin of the collective flow effect in small systems. One school of the theoretical models assumes no contribution of deconfined parton matter evolution in small systems and employs either initial state correlations~\cite{Schenke:2014zha,Schenke:2016lrs,Zhang:2020ayy,Altinoluk:2020psk,Shi:2020djm} or final state interactions among partons~\cite{dEnterria:2010xip}, overlapped strings~\cite{Bierlich:2015rha,Bierlich:2016vgw,Bierlich:2017vhg,Bierlich:2020naj} and cascading hadrons~\cite{Bierlich:2021poz,Zhou:2015iba} to generate the collectivity. The other school of the phenomenological studies incorporate a parton evolution phase dominated by the quark and gluon degrees of freedom, in which hydrodynamic expansions or parton cascade effects lead to the sizable azimuthal anisotropies via the final state interactions in the parton evolution stage~\cite{He:2015hfa,Deng:2011at,Ma:2014pva,Zhao:2017rgg,Zhao:2020pty,Zhao:2022ayk,Zhao:2021bef,Greif:2017bnr,YuanyuanWang:2023meu}. With all the hot debates on the different mechanisms to induce the collective flow in small systems, it is extremely important to further investigate the collective flow in terms of its multi-particle correlations and \pt dependence in a systematic way. In this work, we focus on the understanding of the flow development in high energy pp collisions. 

A multiphase transport model (AMPT) has been widely used to describe key features of the heavy ion collisions based on the kinetic transport approach~\cite{Lin:2004en,Lin:2021mdn}. With the event by event fluctuating initial conditions, parton rescatterings, coalescence hadronization and the hadron cascades, the string melting version of the AMPT model is expected to reasonably reproduce the collective flow in a wide range of collision systems at different energies~\cite{Zhang:2019utb,Zhang:2021vvp,Zhao:2021bef,Zhang:2022fum,Ji:2023eqn}. The traditional AMPT model employs HIJING~\cite{Wang:1991hta} to generate the initial conditions. Recently, the AMPT model with PYTHIA8 based initial conditions~\cite{Bierlich:2022pfr,Bierlich:2018xfw} is found to describe the multiplicity dependence of the identified hadron \pt spectra in high energy pp collisions at the LHC. By incorporating the sub-nucleon geometric fluctuations for the colliding protons using the constituent quark assumption, which induces a large initial spatial eccentricity, a significant long range correlation is found in the simulation for high multiplicity proton proton collisions~\cite{Zheng:2021jrr}. With the capability to simultaneously describe the $p_T$ spectra and multi-particle correlations in pp collisions, one can exploit the interplay of parton and hadron final state interactions by considering the microscopic dynamical process of the evolving system.   

In this work, we will systematically investigate the elliptic flow coefficient generated in pp collisions at $\sqrt{s}=13$ TeV through the Q-cumulant method relying on the PYTHIA8 based AMPT model. In order to demonstrate the unique features of final state parton evolutions embedded in AMPT, we also explore the collective flow generated solely with the anisotropic parton escape mechanism by performing a random $\phi$ test to remove the hydrodynamic type flow components. The study carried out in this work is expected to help us disentangle the various contributions to the collective flow in high energy pp collisions from the perspective of a transport model. 

The rest of this paper is organized as follows: we elaborate the 
transport model approach for pp collisions and the method to calculate the collective flow in Sec.~\ref{sec:model}. Comparisons of the model calculations to the elliptic flow data from pp collisions at the LHC energy are made in Sec.~\ref{sec:results}. Impacts of the parton escape mechanism are explored in Sec.~\ref{sec:discussion}. Finally, we summarize our major conclusions in Sec.~\ref{sec:summary}.

\section{The model and analysis method}
\label{sec:model}

The calculations of pp collisions at $\sqrt{s}=13$ TeV in this work are based on the string melting AMPT model with event-by-event fluctuating initial conditions generated with the PYTHIA/Angantyr framework~\cite{Bierlich:2022pfr,Bierlich:2018xfw}. Sub-nucleon spatial geometries are introduced by dividing a proton into three constituent quarks to deliver the initial spatial eccentricities in the transverse plane. The space-time configuration of the initial parton system is determined by assigning the parton sources to the binary collision center of each colliding constituent quark pair from two proton beams. The sub-nucleon geometry and its strong spatial fluctuations can be important to interpret the collective flow in small systems~\cite{Schenke:2021mxx,Schenke:2014zha,Mantysaari:2016ykx,Welsh:2016siu,Mantysaari:2017cni}. Partons created in the string melting process are then allowed to have further parton rescatterings based on the kinetic transport approach implemented within the Zhang's Parton Cascade (ZPC) model~\cite{Zhang:1997ej}. The parton rescattering cross section in ZPC is set to 0.15mb in this work to reproduce the charged hadron elliptic flow data as will be shown in Sect.~\ref{sec:results}. For the partons after the parton cascade process, they will form different hadronic states by combining with their nearest quark partners following the spatial coalescence model~\cite{He:2017tla}. The hadronic objects produced after the the coalescence hadronization process will be transported for the hadron rescattering stage using an extended relativistic transport model (ART)~\cite{Li:1995pra}.


In this work, we calculate the azimuthal flow $v_{n}$ harmonics from multi-particle correlations via the Q-cumulant method~\cite{Bilandzic:2010jr,Bilandzic:2013kga}, which has been widely applied in the experimental analysis at RHIC and LHC. In this method, by employing the Q-vector defined as
\begin{equation}
Q_n=\sum_{i=1}^{M}e^{in\phi_i},
\end{equation}
the averaged two- and four-particle azimuthal correlations for all Reference Flow Particles (RFPs) in a single event can be evaluated 
\begin{eqnarray}
\langle 2 \rangle & = & \frac{|Q_n|^2-M}{M(M-1)}, \label{eqn:avg2} \\ 
\langle 4 \rangle & = & \frac{|Q_n|^4+|Q_{2n}|^2-2\cdot Re[Q_{2n}Q_n^* Q_n^*]}{M(M-1)(M-2)(M-3)} \\ \nonumber
  & & -2\frac{2(M-2)\cdot |Q_n|^2-M(M-3)}{M(M-1)(M-2)},
\end{eqnarray}
where $M$ represents the number of RFPs and $\phi$ is the azimuthal angle of the particle momentum. The corresponding cumulants are thus calculated by
\begin{eqnarray}
c_{n}\{2\} & = & \langle\langle 2 \rangle\rangle, \\
c_{n}\{4\} & = & \langle\langle 4 \rangle\rangle - 2\cdot\langle\langle 2 \rangle\rangle^2,
\end{eqnarray}
with $\langle\langle \rangle\rangle$ denoting the average for all particles over all events.
In order to suppress the non-flow effects to the above two-particle cumulants, the sub-event method with a sizable pseudorapidity gap has been applied to the analysis. The event in this method is divided into two sub-events, A and B, separated by the pseudorapidity gap $|\deta|$. Therefore, one can change Eq.~\ref{eqn:avg2} to
\begin{equation}
\langle 2 \rangle_{\deta} = \frac{Q_n^A\cdot Q_n^{B*}}{M_A\cdot M_B},
\label{eqn:avg2gap}
\end{equation}
in which $Q_n^A$ and $Q_n^B$ are the flow vectors for two sub-events, each having $M_A$ and $M_B$ particles. The two-particle cumulant with $|\deta|$ gap can be straightforwardly obtained by averaging Eq.~\ref{eqn:avg2gap} for all events 
\begin{equation}
c_{n}\{2,|\deta|\} = \langle\langle 2 \rangle\rangle_{\deta}.\label{eqn:cumugap}
\end{equation}
To calculate the differential flow of the Particles Of Interests (POIs), we define the flow vector $p_n$ for the specific hadron species within certain kinematic range
\begin{equation}
p_n=\sum_{i=1}^{m_p}e^{in\phi_i},
\end{equation} 
where $m_p$ represents the number of POIs in an event. After applying the long range $\deta$ gap, the differential two-particle azimuthal cumulant in an event is thus 
\begin{equation}
\langle 2^{\prime}\rangle _{\deta}=\frac{p_{n,A}Q_{n,B}^{*}}{m_{p,A}M_{B}}, \label{eqn:avg2p}
\end{equation} 
where POIs and RFPs involved in the calculation are always in two different sub-event regions, A and B, to remove the non-flow contributions. After defining the differential two-particle cumulant as the average of all events for Eq.~\ref{eqn:avg2p}:
\begin{equation}
d_n\{2,|\deta|\}=\langle\langle 2^{\prime}\rangle\rangle_{\deta},
\end{equation}
one can arrive at the final differential flow $v_{n}$ with pseudorapidity gap as follows
\begin{equation}
v_n\{2,|\deta|\}=\frac{d_n\{2,|\deta|\}}{\sqrt{c_{n}\{2,|\deta|\}}}.
\end{equation}
For the flow estimation in pp collisions, we need to further subtract the non-flow contribution extracted from the low multiplicity events based on the multiplicity scaling assumption. Considering the non-flow effects between two-particle correlations is usually proportional to $1/M$~\cite{Zhou:2015iba}, a method of subtracting these non-flow contributions is commonly used in experiments~\cite{ALICE:2022ruh,ALICE:2015lpx,CMS:2016fnw} following the multiplicity scaling assumption. With the multiplicity ratio $k=\langle M\rangle ^{low} /\langle M\rangle$, the two-particle azimuthal cumulants in different multiplicity bins with the average particle number $\langle M\rangle$ after non-flow subtraction can be obtained as
\begin{equation}
c_n^{sub}\{2,|\deta|\}=c_{n}\{2,|\deta|\}-k\cdot c_{n}^{low}\{2,|\deta|\}. 
\label{eqn:cnsub}
\end{equation}
The superscript notation $low$ denotes the corresponding quantities estimated from the low multiplicity events. The non-flow subtracted azimuthal flow can thus be defined accordingly with
\begin{equation}
v_n^{sub}\{2,|\deta|\}=\frac{d_n\{2,|\deta|\}-k\cdot d_n^{low}\{2,|\deta|\} }{\sqrt{c_{n}\{2,|\deta|\}-k\cdot c_{n}^{low}\{2,|\deta|\}}}. 
\label{eqn:vnsub}
\end{equation}
 To suppress the event level multiplicity fluctuation effects, we follow the strategy used in~\cite{Jia:2017hbm,Zhao:2021bef,ATLAS:2017hap}, which performs the flow analysis in each selected reference particle number $M$ bin and then map each $M$ interval into the event activity estimator on the basis of selected charged particle number $N_{ch}$ with $p_{T}>0.4$ GeV$/c$. In the rest of this work, the flow analysis is carried out with the reference particles defined using the charged hadrons within $|\eta|<2.4$ and $0.3<p_{T}<3$ GeV$/c$. The low multiplicity events used for the non-flow subtraction are required to have reference particle number $10<M<20$.

\begin{figure*}[hbt!]
	\centering
	\includegraphics[width=0.49\textwidth]{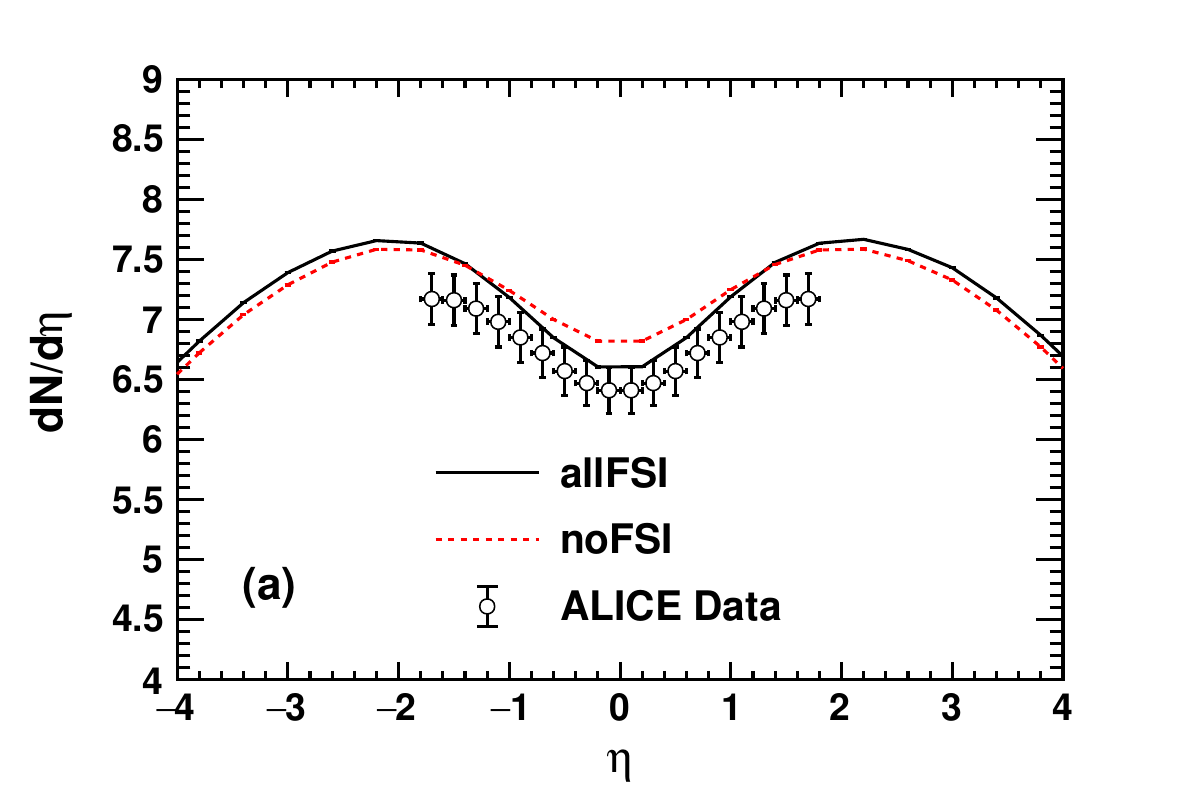}
	\includegraphics[width=0.49\textwidth]{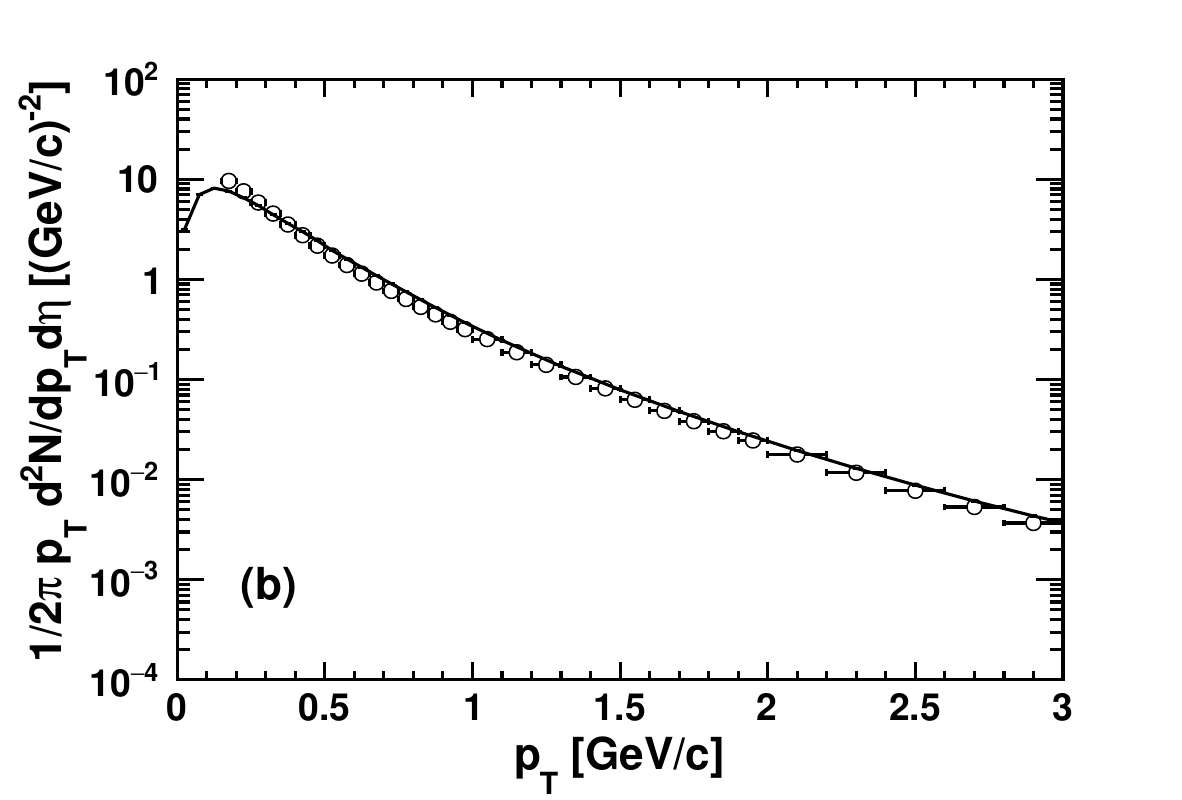}
	\caption{(Colour Online) The pseudo-rapidity distribution (a) and the transverse momentum spectra (b) for charged hadrons in pp collisions at $\sqrt{s}=13$ TeV from the AMPT model calculations with all final state parton and hadron rescatterings (allFSI) and without final state interactions (noFSI) compared to ALICE data with at least one charged hadron in the mid-rapidity region (INEL$>$0 events). The allFSI and noFSI (only $dN/d\eta$) results are represented by the solid lines and dashed lines, respectively. ALICE data indicated by the open circles are taken from Ref.~\cite{Adam:2015pza}}
	\label{fig:charge_dndeta}       
\end{figure*}
\begin{figure}[hbt!]
	\centering
	\includegraphics[width=0.5\textwidth]{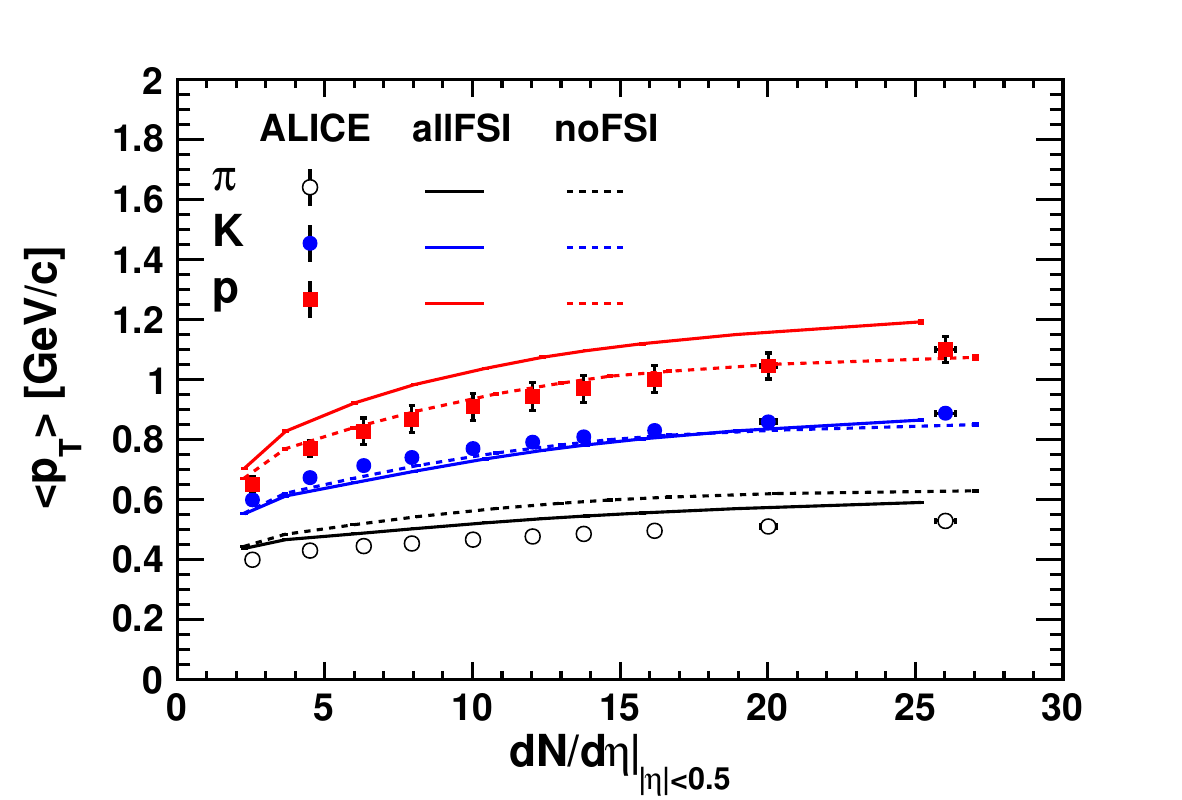}
	\caption{(Colour Online) Average transverse momentum for $\pi$, K and proton within $|y|<0.5$ and $0<\pt<3$ GeV$/c$ varying with the charged particle density from the AMPT model calculations with all final state parton and hadron rescatterings (allFSI) and without final state interactions (noFSI) compared to ALICE data in pp collisions at $\sqrt{s}=13$ TeV. The allFSI and noFSI results are represented by the solid lines and dashed lines, respectively. ALICE data indicated by the markers are taken from Ref.~\cite{Acharya:2020zji}.}
	\label{fig:avgpt_pikp_cent}    
\end{figure}

\section{Results and discussions}
\label{sec:results}

\subsection{Hadron productions}
To begin with, we compare the charged  particle productions in pp collisions at $\sqrt{s}=13$ TeV of the AMPT model calculations with PYTHIA8 initial conditions to the experimental data. The results of the model calculations shown by the solid lines are obtained when all the final state parton/hadron rescatterings and resonance decays are finished. Events with at least one charged track in mid-rapidity $|\eta|<1$ are selected for the comparison (INEL$>$0 events). It is found in Fig.~\ref{fig:charge_dndeta}(a) that the charged particle density distributions per unit pseudorapidity space measured in the experiments~\cite{Adam:2015pza} are well described by the model calculations. We also present the results of charged hadron transverse momentum spectra in Fig.~\ref{fig:charge_dndeta}(b). The model calculations are found to reproduce the experimental data of transverse momentum spectra in a reasonable way. We also present the results without any final state interactions in Fig.~\ref{fig:charge_dndeta} shown by the red dashed line. The results without any final state interactions are supposed to be similar to the pure PYTHIA8 predictions. It can be found in this comparison that the final state interactions and coalescence hadronization mechanism incorporated in the AMPT framework will slightly reduce the multiplicity number at final state for high energy pp collisions.

We explore the event activity dependence of identified hadron productions by examining the average transverse momentum $\langle\pt\rangle$ for $\pi^{\pm}$, $K^{\pm}$, $p$ and $\bar{p}$ versus the charged particle multiplicity in Fig.~\ref{fig:avgpt_pikp_cent}. The identified particles within $|y|<0.5$ having $0<\pt<3$ GeV$/c$ are used in this analysis. The results are obtained following the experimental procedure in Ref.~\cite{Acharya:2020zji}. In order to restrict the comparison to the low \pt regime where the string melting version of the AMPT model is more applicable, the experimental data are extracted by refitting the measured \pt spectra using the Tsallis distribution in every event class within $0<\pt<3$ GeV$/c$. The experimentally observed strong increase of average \pt with the increasing event multiplicity is reproduced in the model calculations. The mass hierarchy of $\langle\pt\rangle$, which is connected to the radial flow effect, is found to be naturally included in this framework. The results with and without final state interactions are presented by the solid and dashed lines in this figure. The observation is consistent with our previous study~\cite{Zheng:2021jrr} that the multiplicity dependence of the average \pt for different hadrons arises from the convolution of multiple parton interactions implemented in the PYTHIA8 initial conditions. The final state interactions further enlarge the mass hierarchy of the \pt distributions for different particle species and result in the slight overestimation of the average \pt for protons. Compared to our previous results in Ref.~\cite{Zheng:2021jrr}, the parton scattering cross section has been tuned from 0.2 mb to 0.15mb in this work to have a better agreement with the elliptic flow data observed in pp collisions at $\sqrt{s}=13$ TeV. It is also seen that the hadron yield and transverse momentum distributions are generally not sensitive to the parton scattering cross sections.

\begin{figure*}[hbt!]
	\centering
	\includegraphics[width=0.49\textwidth]{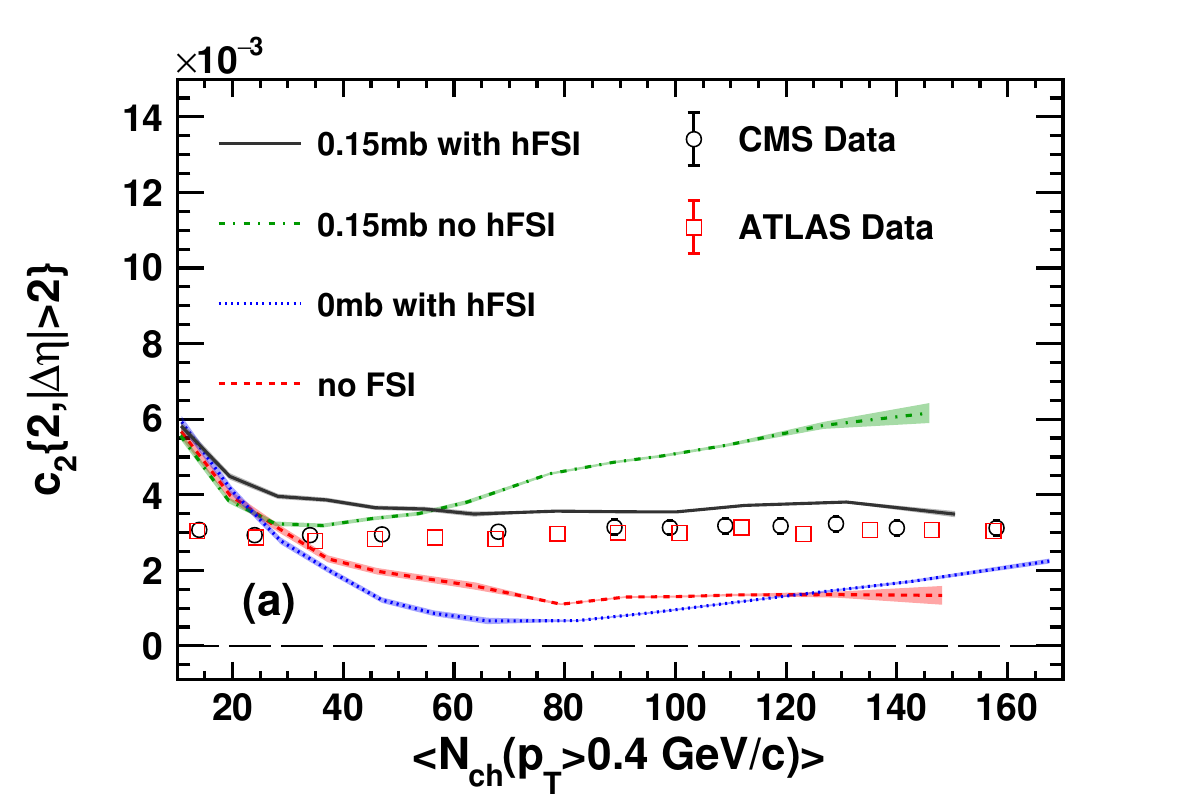}
	\includegraphics[width=0.49\textwidth]{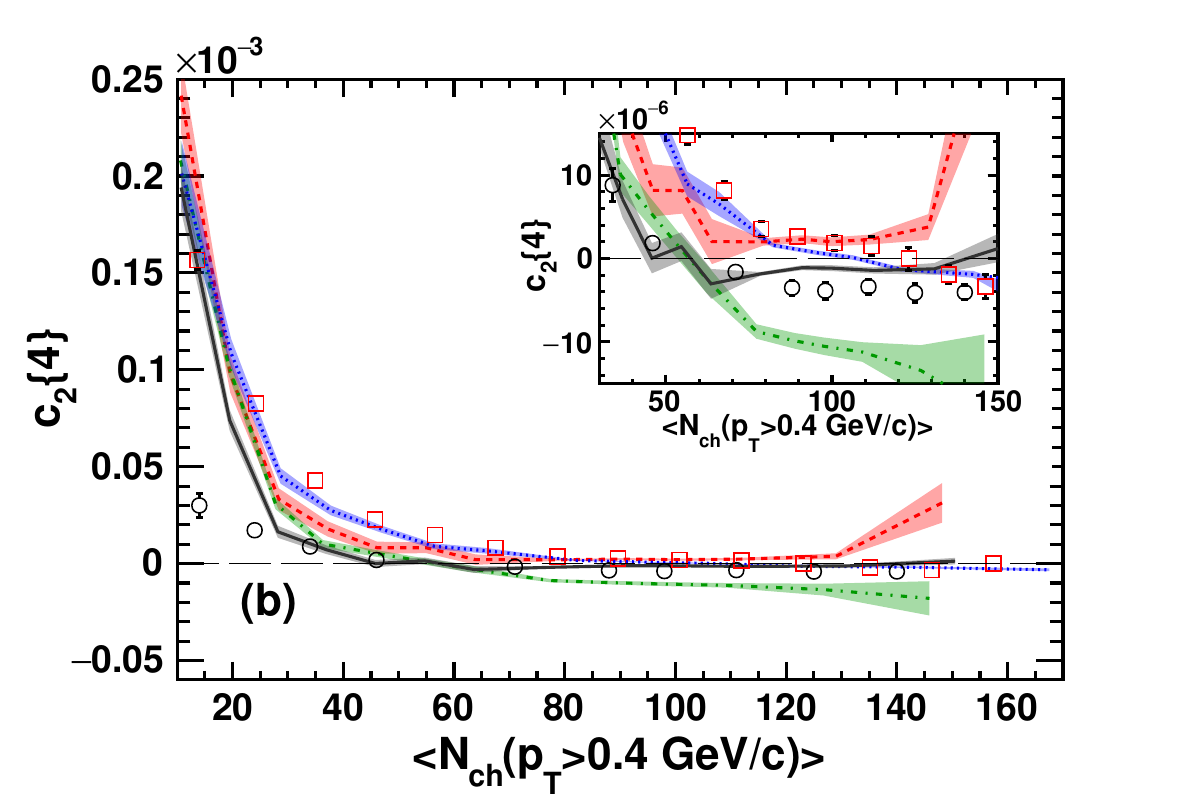}
	\caption{(Colour Online) Two-particle cumulants in the sub-event method (a) and four-particle cumulants (b) for charged hadrons with $|\eta|<2.4$ and $0.3<\pt<3$ GeV$/c$ in pp collisions at $\sqrt{s}=13$ TeV versus the average selected charged hadron number $\langle N_{ch}\rangle$ from the model calculations and the experimental data. The gray solid lines represent the calculations with all final state interactions (0.15mb with hFSI). The results with only final state parton rescatterings (0.15mb no hFSI), only final state hadron rescatterings (0mb with hFSI) and no final state interactions (no FSI) are shown by the green dash-dotted lines, blue dotted lines and red dashed lines, respectively. The bands represent the statistical uncertainties. 
	CMS data shown by the open circles are taken from Ref.~\cite{CMS:2016fnw}. ATLAS data shown by the open squares are taken from Ref.~\cite{ATLAS:2017hap}. }
	\label{fig:c2_charge_2and4}    
\end{figure*}

\begin{figure*}[hbt!]
	\centering
	\includegraphics[width=0.49\textwidth]{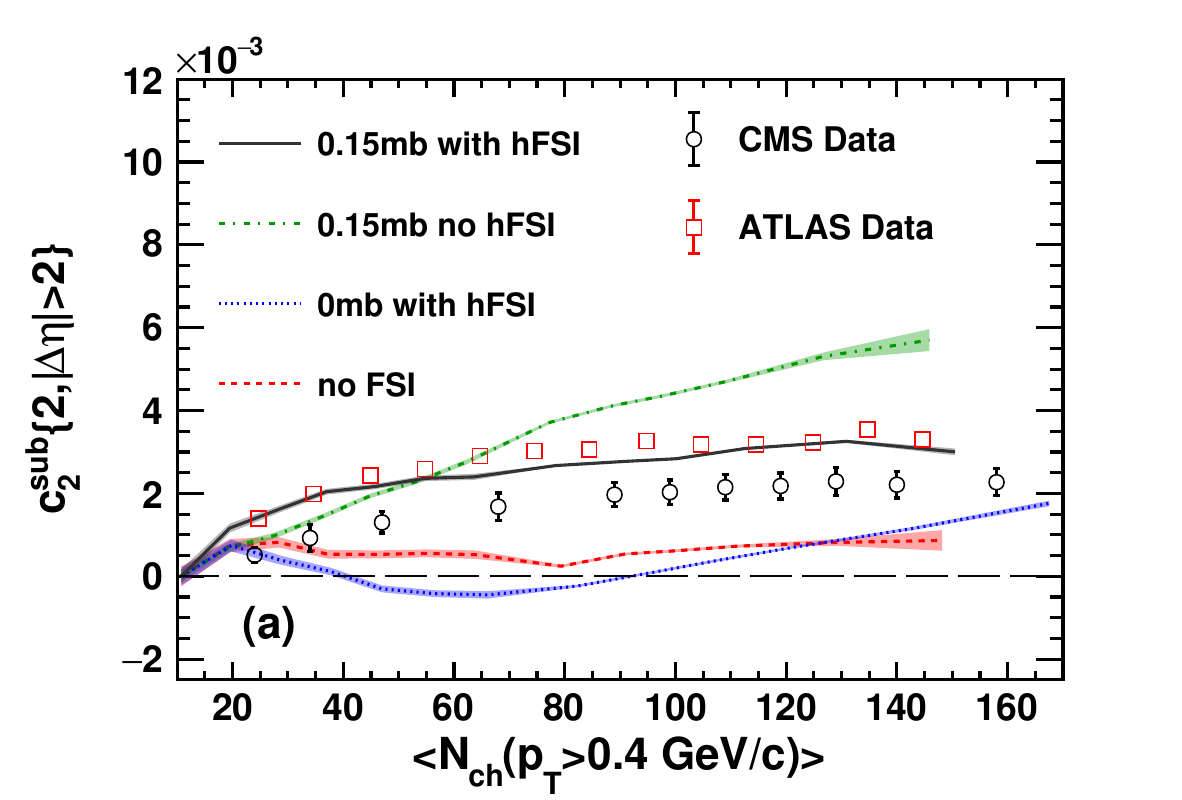}
	\includegraphics[width=0.49\textwidth]{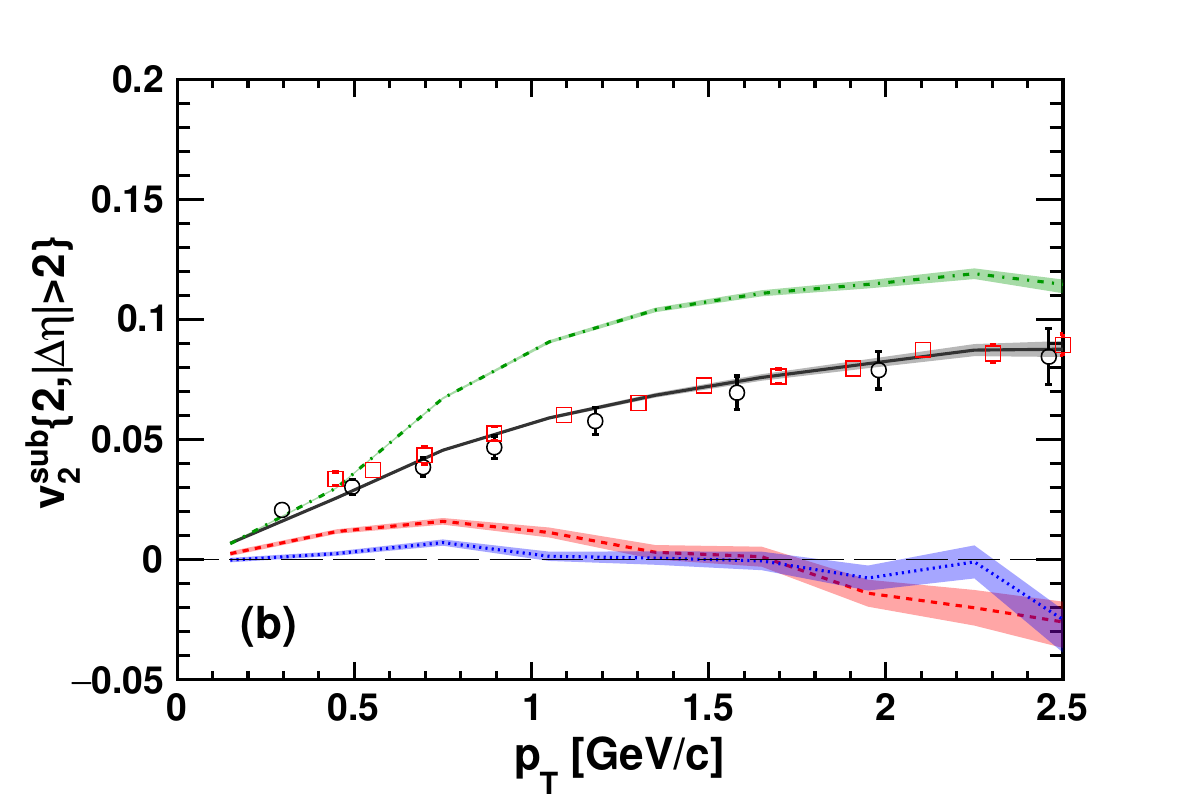}
	\caption{(Colour Online) Two-particle cumulant (a) and extracted \vtwo coefficients in high multiplicity events with $M>100$ (b) based on the sub-event method using $|\Delta\eta|>2$ after subtracting the non-flow contribution with low multiplicity events for charged hadrons in pp collisions at $\sqrt{s}=13$ TeV. Charged hadrons with $|\eta|<2.4$ and $0.3<\pt<3$ GeV$/c$ are defined as reference flow particles in this analysis. The gray solid lines represent the calculations with all final state interactions (0.15mb with hFSI). The results with only final state parton rescatterings (0.15mb no hFSI), only final state hadron rescatterings (0mb with hFSI) and no final state interactions (no FSI) are shown by the green dash-dotted lines, blue dotted lines and red dashed lines, respectively. The bands represent the statistical uncertainties. CMS data shown by the open circles are taken from Ref.~\cite{CMS:2016fnw}. ATLAS data shown by the open squares are taken from Ref.~\cite{ATLAS:2016yzd}.
	}
	\label{fig:c2v2_charge_sub}    
\end{figure*}

\begin{figure*}[bp]
	\centering
	\includegraphics[width=0.43\textwidth]{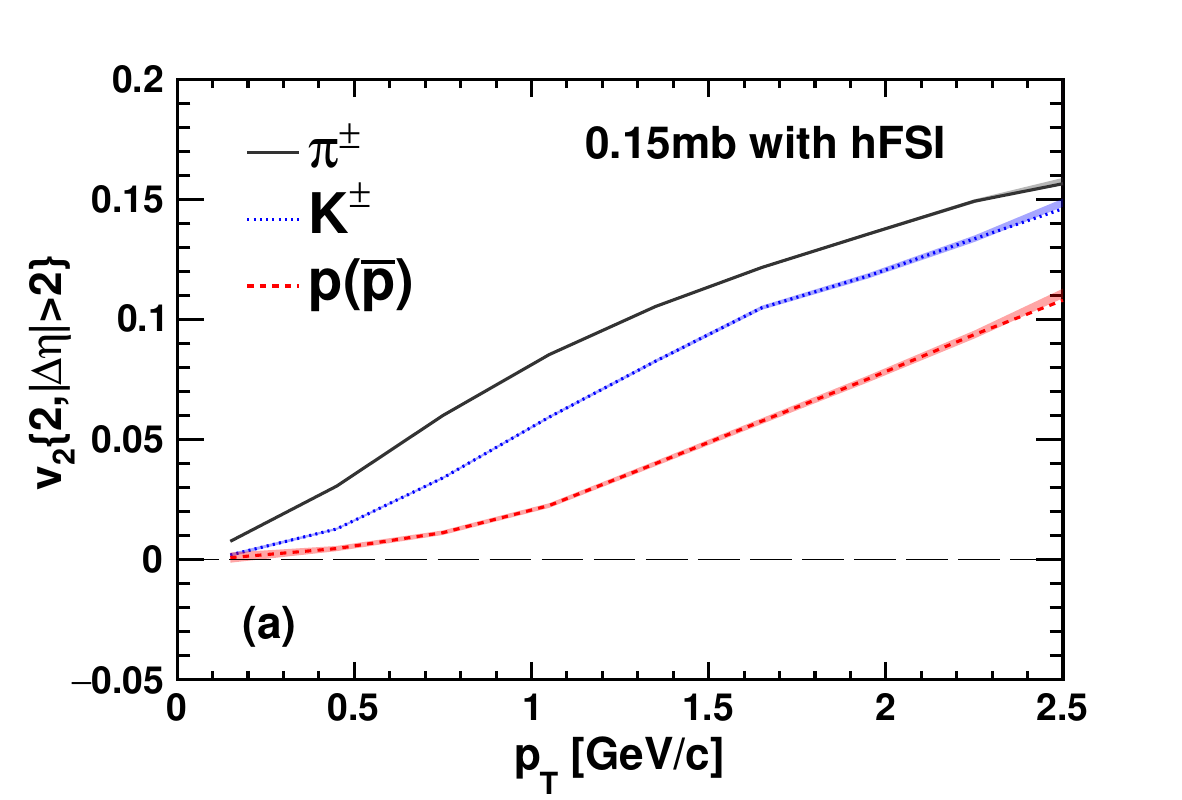}
	\includegraphics[width=0.43\textwidth]{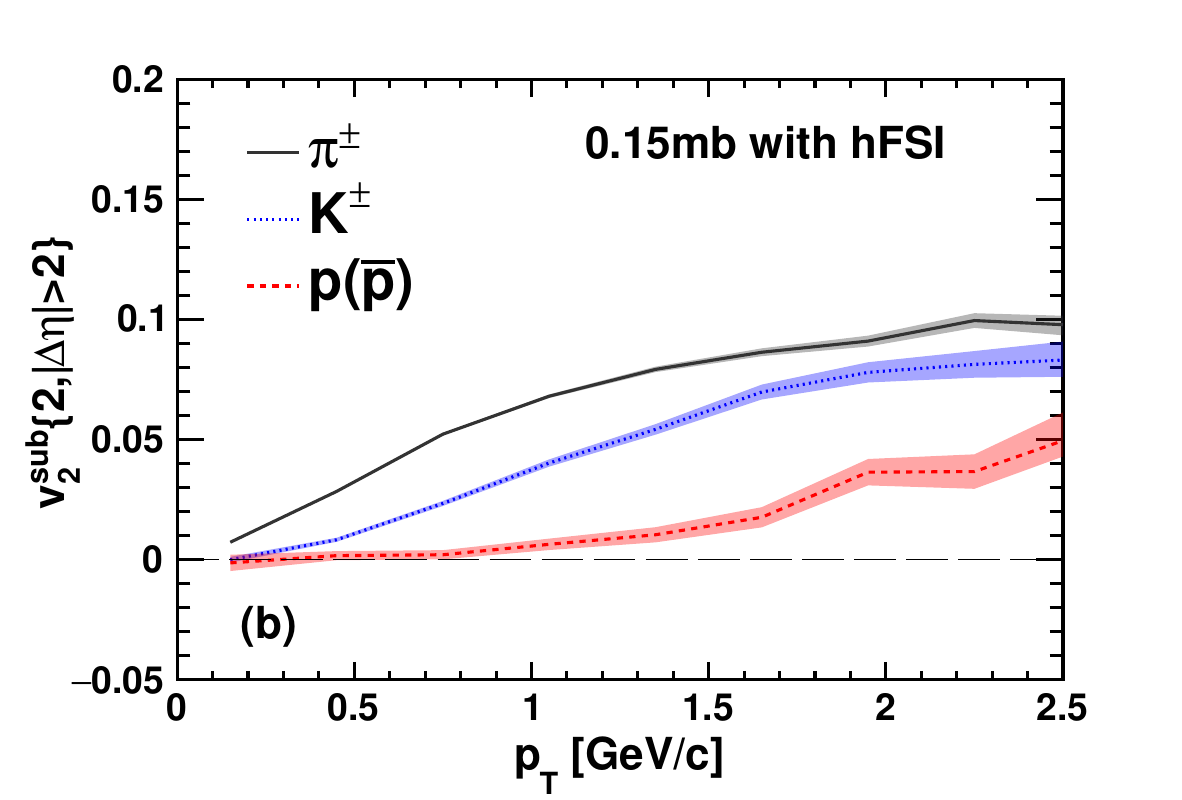}
	\includegraphics[width=0.43\textwidth]{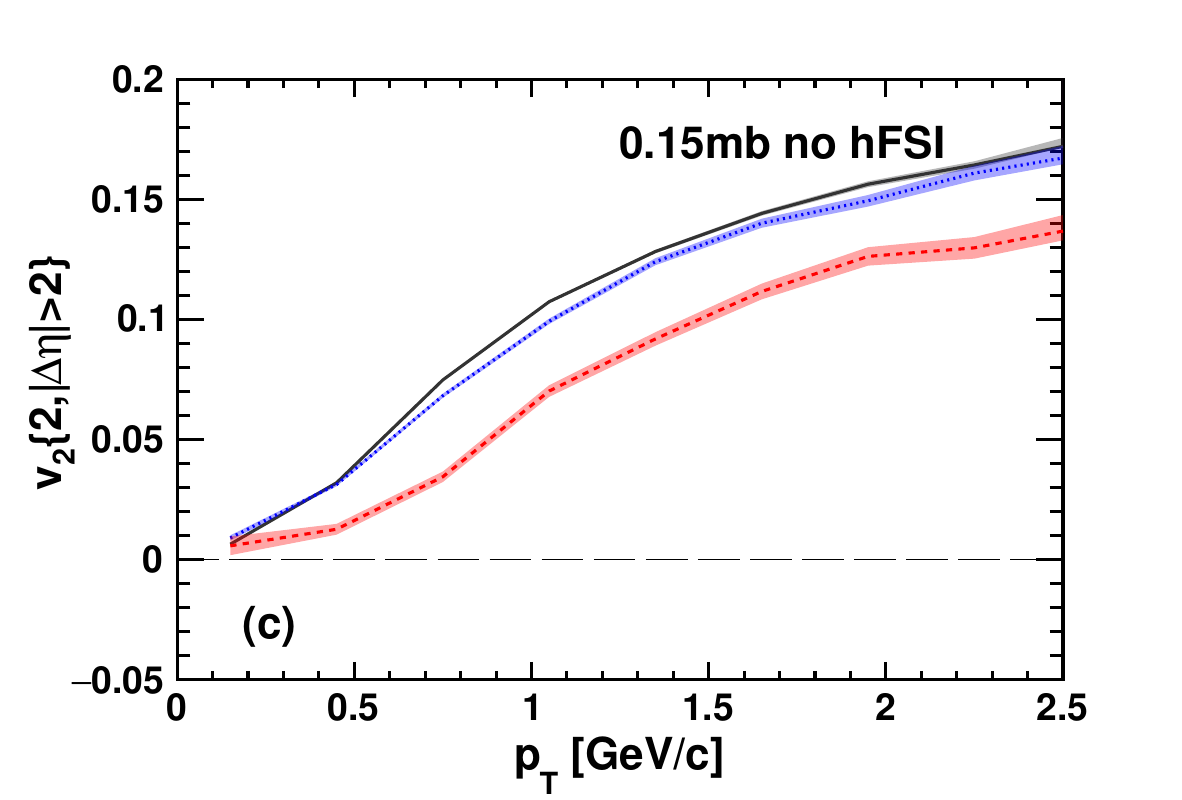}
	\includegraphics[width=0.43\textwidth]{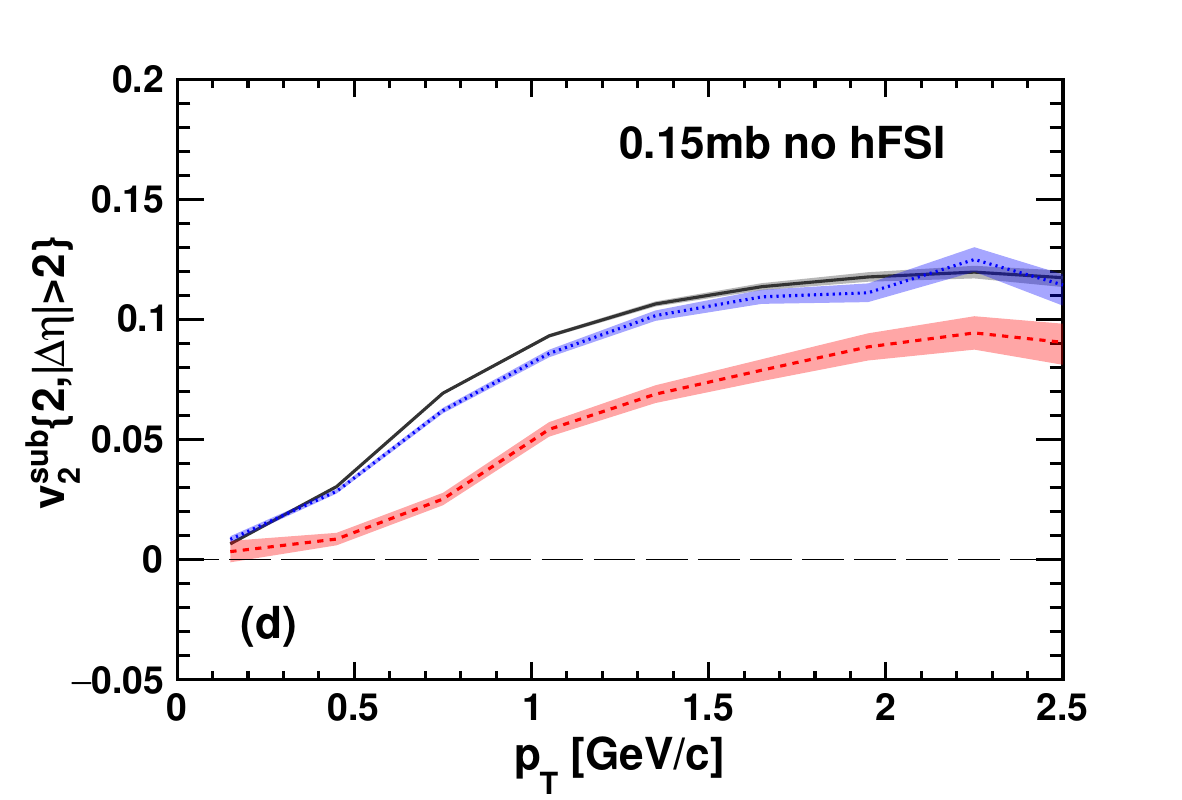}
	\includegraphics[width=0.43\textwidth]{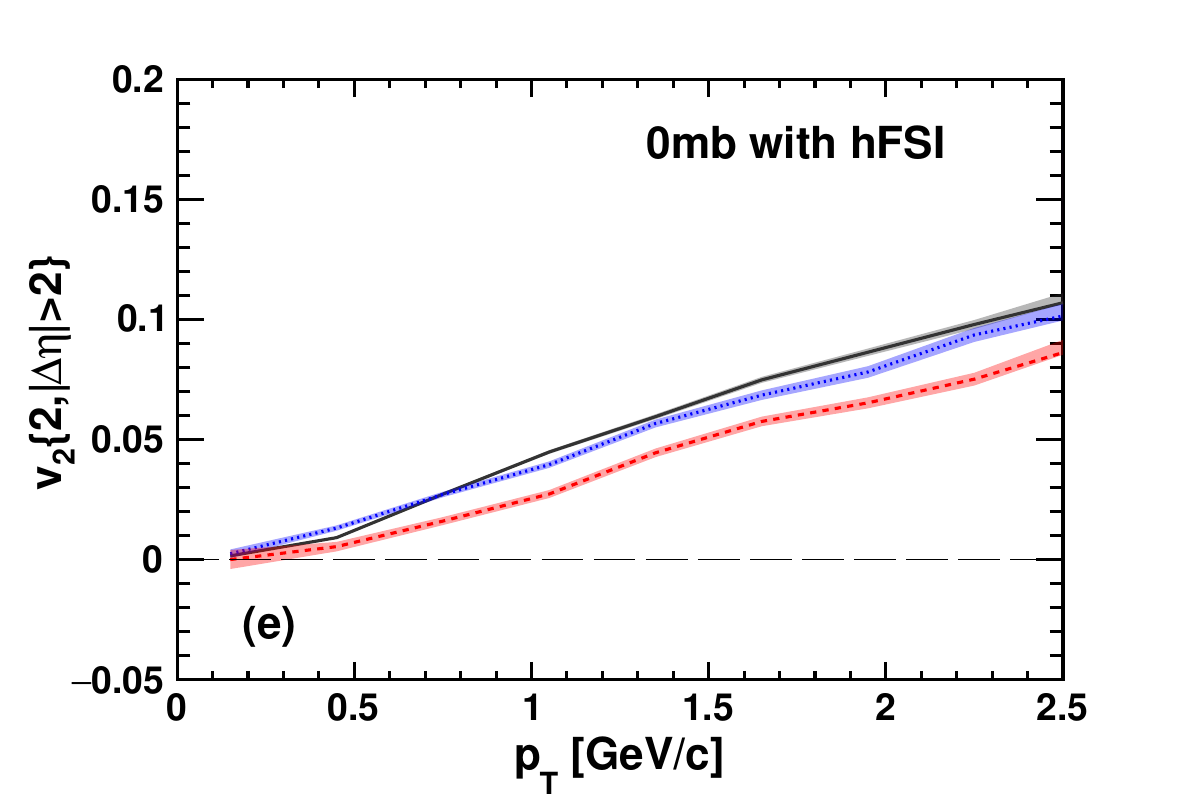}
	\includegraphics[width=0.43\textwidth]{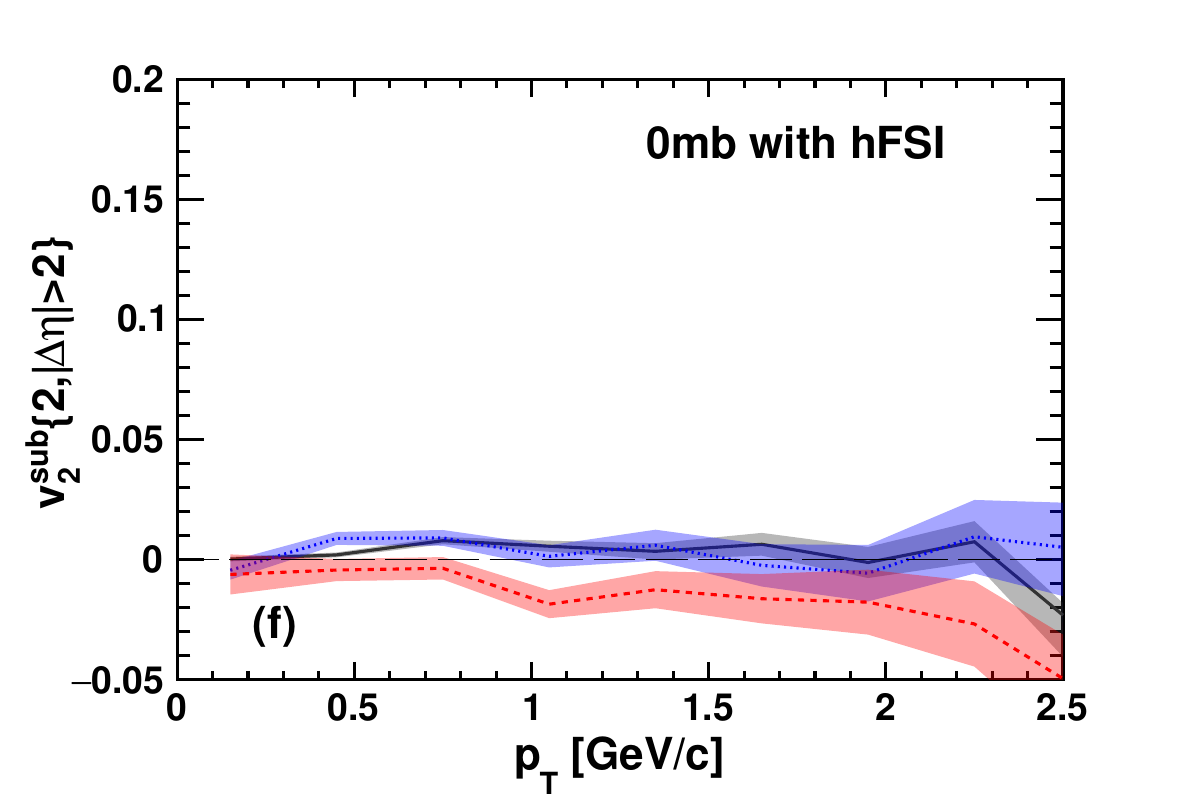}
	\includegraphics[width=0.43\textwidth]{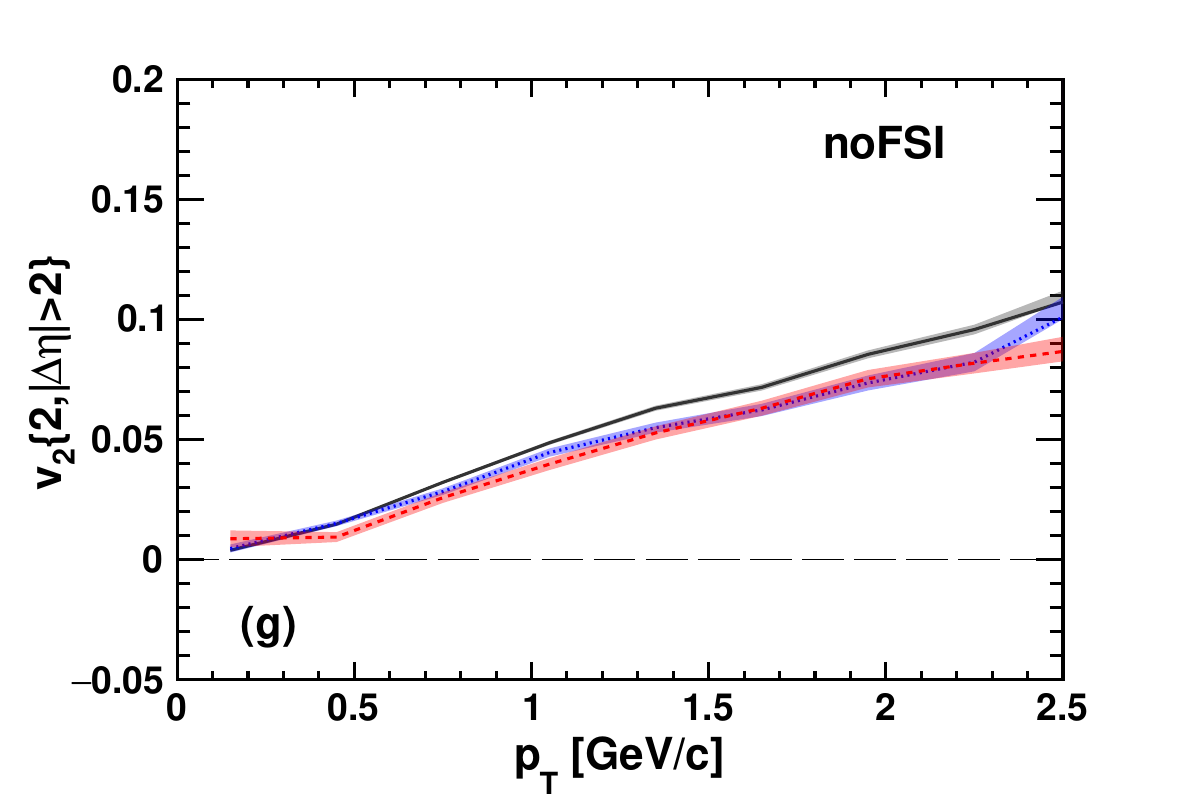}
	\includegraphics[width=0.43\textwidth]{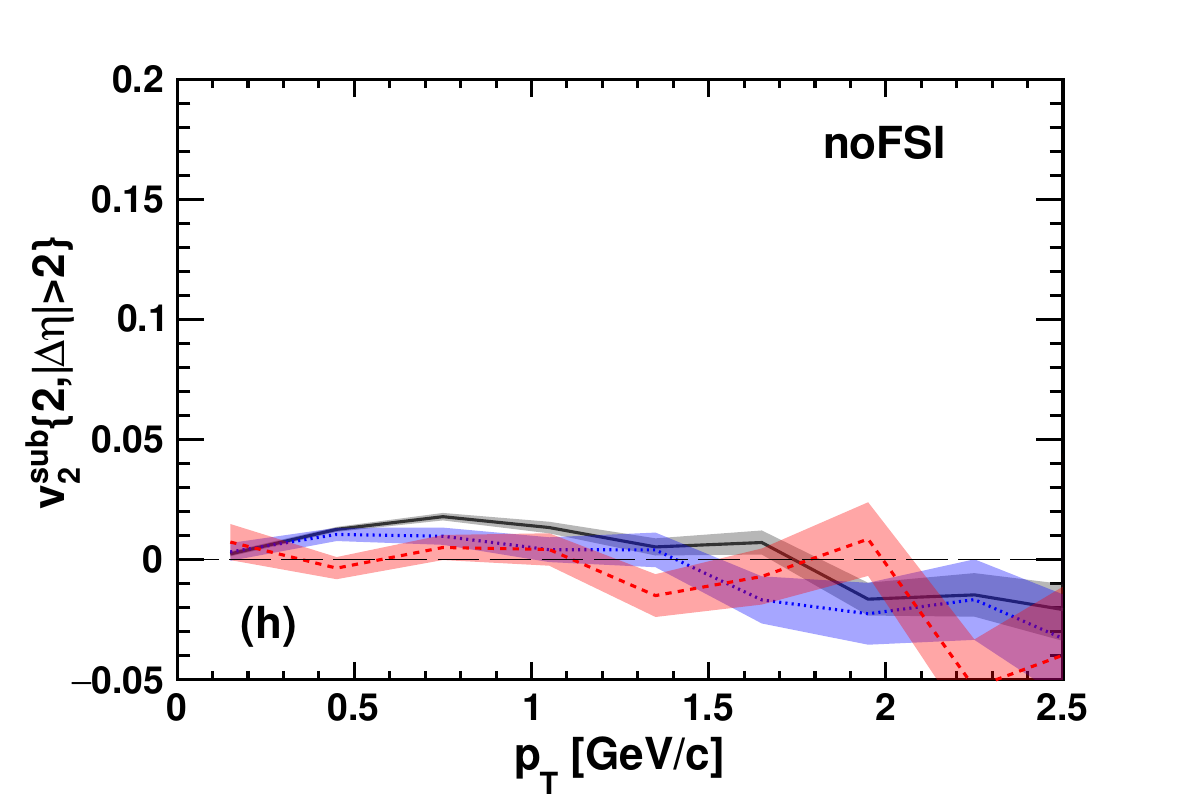}
	\caption{(Colour Online) The \pt differential \vtwo (left column) and $\vtwo^{sub}$ (right column) of identified hadrons based on the sub-event method using pseudorapidity gap $|\deta|>2$ for high multiplicity events with $M>100$ in pp collisions at $\sqrt{s}=13$ TeV. The AMPT results with all the final state interactions (0.15mb with hFSI), only parton rescatterings (0.15mb no hFSI), only hadronic rescatterings (0mb with hFSI) and no final state interactions (noFSI) are shown in the first to the fourth row, respectively. The $\pi^{\pm}$, $K^{\pm}$, and proton results are shown by gray solid lines, blue dotted lines and red dashed lines, respectively. The bands represent the statistical uncertainties.}
	\label{fig:v2_piKp}    
\end{figure*}

\subsection{Azimuthal flow with Q-cumulant method}
As a next step, we investigate the collective flow coefficient in pp collisions at $\sqrt{s}$=13 TeV using particles within the pseudorapidity acceptance range $|\eta|<2.4$. All charged particles with $|\eta|<2.4$ and $0.3<\pt<3$ GeV$/c$ are used as reference flow particles in this study. The number of charged particles $N_{ch}$ selected with $|\eta|<2.4$ and $\pt>0.4$ GeV$/c$ is used to estimate the event activity. The low multiplicity events, with which we extract the non-flow contributions, are required to have $10<M<20$. A pseudorapidity gap $|\deta|>2$ has been applied to obtain the two-particle cumulants in the flow calculations using the sub-event method. To estimate the scaling factor $k$ used in Eq.~\ref{eqn:cnsub} and Eq.~\ref{eqn:vnsub}, the average number of reference flow particles $\langle M\rangle$ in each multiplicity interval has been used. The event averaging is performed in each reference particle number interval and then mapped to $\langle N_{ch}\rangle$ in order to remove the multiplicity fluctuation effects~\cite{Jia:2017hbm}.

We present the two-particle cumulants $\ctwo$ after applying $\deta$ cut with the sub-event method and four-particle cumulants $\cfour$ for charged hadrons in Fig.~\ref{fig:c2_charge_2and4}. The impacts of final state parton and hadron evolution stages on the building of multi-particle cumulants are explored by switching on each key ingredient implemented in AMPT step by step. In Fig.~\ref{fig:c2_charge_2and4}(a), the black open circles and red open squares represent the $\ctwo$ experimental data with $|\Delta\eta|>2$ measured by CMS~\cite{CMS:2016fnw} and ATLAS~\cite{ATLAS:2017hap}, respectively. The red dashed line shows the results from AMPT calculations without any final state parton or hadron rescatterings. The green dash-dotted line and the blue dotted line represents the scenario with only partonic interactions and hadronic interactions, respectively. The result with all the final state interactions is shown by the gray solid line. The bands represent the statistical uncertainties of the model calculations. It is found that with all the final state parton and hadron interactions included, the model calculation describes the observed charged two-particle cumulants reasonably, except at $\langle N_{ch}\rangle$ below 30. Comparing the partonic and hadronic final state interactions, the sizable $\ctwo$ cumulants appear mainly during the parton evolution stage and the inclusion of additional hadronic rescattering generally diminishes the two particle correlations. The reduction of $\ctwo$ after incorporating hadronic final state interactions might be related to the rapid isotropization of the parton matter created in pp collisions. The sizable initial eccentricity induced through sub-nucleon spatial fluctuations can drop rapidly during the expansion of the parton system. The follow-up hadronic scatterings will decrease the momentum space anisotropy with a more isotropized spatial configuration. 
With only hadronic rescattering effect, the result is quite close to the case without any final state interactions, compatible with the long range correlation studies~\cite{Zheng:2021jrr}. It is also shown that the results either with or without final state interactions become quite similar in the region of $\langle N_{ch}\rangle$ less than 20, indicating non-flow effects are dominant in the low multiplicity region. 

Multi-particle cumulant $\cfour$ is supposed to be less sensitive to the non-flow effects and uncover the collective motion of the particle system in a more exclusive way. It is expected for AA collisions that $\cfour$ should be negative if the correlation comes from the hydrodynamic collective flow~\cite{Bzdak:2013rya,Zhao:2021bef,Jia:2017hbm}. In Fig.~\ref{fig:c2_charge_2and4}(b), we show the comparisons for the four-particle cumulants $\cfour$ varying with the event multiplicity represented by average charged particle number $\langle N_{ch}\rangle$. The discrepancy between ATLAS and CMS data arises due to the treatment on the selection of the events. Events are analyzed in reference particle number bins and then mapped to corresponding $N_{ch}$ values in ATLAS data while event averaging is performed directly with $N_{ch}$ in CMS data~\cite{ATLAS:2017hap}. We find that the observed negative $\cfour$ in high multiplicity events from AMPT is predominantly coming from the parton evolution stage similar to the arise of the sizable two-particle cumulant as seen in Fig.~\ref{fig:c2_charge_2and4}(a). This is similar to the finding of Ref.~\cite{Zhao:2024wqs}, which shows that the partonic interactions are essential for the flow like correlations in high multiplicity jets. $\cfour$ with only final state parton rescatterings reaches smaller value in the negative region. The inset shows that with only partonic interactions $\cfour$ becomes negative at similar $N_{ch}$ compared to the scenario that both partonic and hadronic final interactions are turned on. With all the final state parton and hadron interactions, the gray line follows the decreasing trend as the increase of the event multiplicity, qualitatively similar to the behavior of ATLAS data. It is also noticed that the gray line turns into negative earlier than the ATLAS data. We also confirmed the event averaging effect with our model calculations. By using an event averaging method based on the number of selected charged particles similar to the CMS analysis procedure, the results with all the final state effects become close to the CMS data, reinstating the discrepancy between ATLAS and CMS data. 

To further investigate the flow coefficients with the two-particle correlation method, we subtract the non-flow contribution using the two-particle cumulants in low multiplicity events from \ctwo in the sub-event method following the prescription in Eq.~\ref{eqn:cnsub}. 
It is observed in Fig.~\ref{fig:c2v2_charge_sub}(a) that the subtracted \ctwo is approaching zero in the low multiplicity region for all model setups, suggesting the impact of non-flow subtraction. However, one can see that the non-flow effect is not completely removed as the red line is not at zero at finite $N_{ch}$. The final \ctwo with both parton and hadron rescatterings is initially growing with $N_{ch}$ and then saturated following the trend of the experimental data. The ATLAS data are slightly higher than the CMS data in Fig.~\ref{fig:c2v2_charge_sub}(a) simply because a higher $p_T$ cut $0.5<\pt<5$ GeV$/c$ has been applied for charged hadron selections in that analysis~\cite{ATLAS:2016yzd}.
It is also found that with only hadronic final state interactions, no collective flow signal can be generated except in the very high multiplicity region.
In Fig.~\ref{fig:c2v2_charge_sub}(b), we investigate the \pt differential charged particle \vtwo after low multiplicity non-flow subtraction for the high multiplicity events with $M>100$. The subtracted \pt differential \vtwo data are found to be well described by the model calculations at final state. $v_{2}^{sub}\{2\}(p_{T})$ is increasing with the increase of \pt and becomes saturated in the higher \pt region. Without the partonic evolution stage (shown by the red dashed lines and the blue dotted lines), the subtracted \vtwo is close to zero across the whole \pt region.

Finally, we present the high multiplicity event $\vtwo(\pt)$ of identified particles in Fig.~\ref{fig:v2_piKp} and emphasize that the mass ordering effect of $\vtwo(\pt)$ can be a clear signature of the existence of a final state parton evolution stage in high energy pp collisions. The extracted \vtwo coefficients without subtracting the low multiplicity event modulations are shown in the left column of Fig.~\ref{fig:v2_piKp}. A clear mass ordering feature is observed in Fig.~\ref{fig:v2_piKp}(a) for the AMPT calculations with all final state parton and hadron interactions switched on, in which heavier particles have smaller elliptic flow with $\vtwo(\pi^{\pm}) > \vtwo(K^{\pm}) > \vtwo(p+\bar{p})$ in the low \pt region. The \vtwo value is found to be larger for all particle species when the hadronic evolution stage is off as shown in Fig.~\ref{fig:v2_piKp}(c). The separation between the \vtwo of various particle species seen in Fig.~\ref{fig:v2_piKp}(a) is significantly reduced, which can be accounted by the different responses of these hadron species in the hadron rescattering process.
We also explore the development of \vtwo for identified hadrons by only switching on the hadron level rescatterings or turning off all the final state interactions shown by Fig.~\ref{fig:v2_piKp} (e) and (g). In Fig.~\ref{fig:v2_piKp}(e), slight mass ordering effects exist due to the hadronic rescatterings, while no significant mass ordering feature can be resolved in Fig.~\ref{fig:v2_piKp}(g). It is inferred in these comparisons that the mass ordering feature is mainly developed in the partonic evolution stages arising from the quark coalescence formalism and further enlarged during the final state hadronic interactions. As is discussed in Ref.~\cite{Li:2016ubw,Li:2016flp}, the mass ordering of \vtwo in the transport model framework can be naturally understood with the kinematics of coalescence process. However, we notice that although hadron rescattering effects are important for mass ordering in large collision systems or dAu collisions at RHIC energy~\cite{Li:2016ubw,Li:2016flp}, pPb collisions at LHC energy scale are less affected by hadronic interactions and more sensitive to the partonic stage evolutions as is shown in Ref.~\cite{Li:2016ubw,Sarkar:2016nox,Tang:2023wcd}, consistent with our observations in high energy pp collisions. 

After implementing the low multiplicity subtraction method deduced in Sect.~\ref{sec:model}, the mass ordering feature of $\vtwo^{sub}$ in each model setup is similar to the $\vtwo$ case as is shown in the right column of Fig.~\ref{fig:v2_piKp}. Although the $\vtwo^{sub}$ magnitude decreases after subtractions, the result indicates that the mass ordering effect uncovered in the comparison is robust either with or without the non-flow contaminations. It is also observed that the subtracted $\vtwo^{sub}$ for all hadron species become close to zero in the case with only hadronic rescatterings or without any final state interactions, making it impossible to identify any mass ordering features in these two scenarios. This observation suggests that the sizable \vtwo shown in Fig.~\ref{fig:v2_piKp} (e) and (g) are therefore dominant by the non-flow contributions. In the comparisons of this section, we find that the final state parton evolution phase and the coalescence hadronization process coupled to it are important ingredients to produce the sizable elliptic flow signal and significant mass ordering features in high energy pp collisions. It has been shown that very few collisions in transport model approach might generate significant flow feature in small systems like pp collisions~\cite{He:2015hfa,Greif:2017bnr,Kurkela:2018ygx,Borghini:2010hy}. The observations in this work can be another example suggesting that the full hydrodynamization process may not be necessary to explain the experimentally observed collectivity in pp collisions~\cite{Grosse-Oetringhaus:2024bwr}.

\begin{figure*}[hbt]
	\centering
	\includegraphics[width=0.49\textwidth]{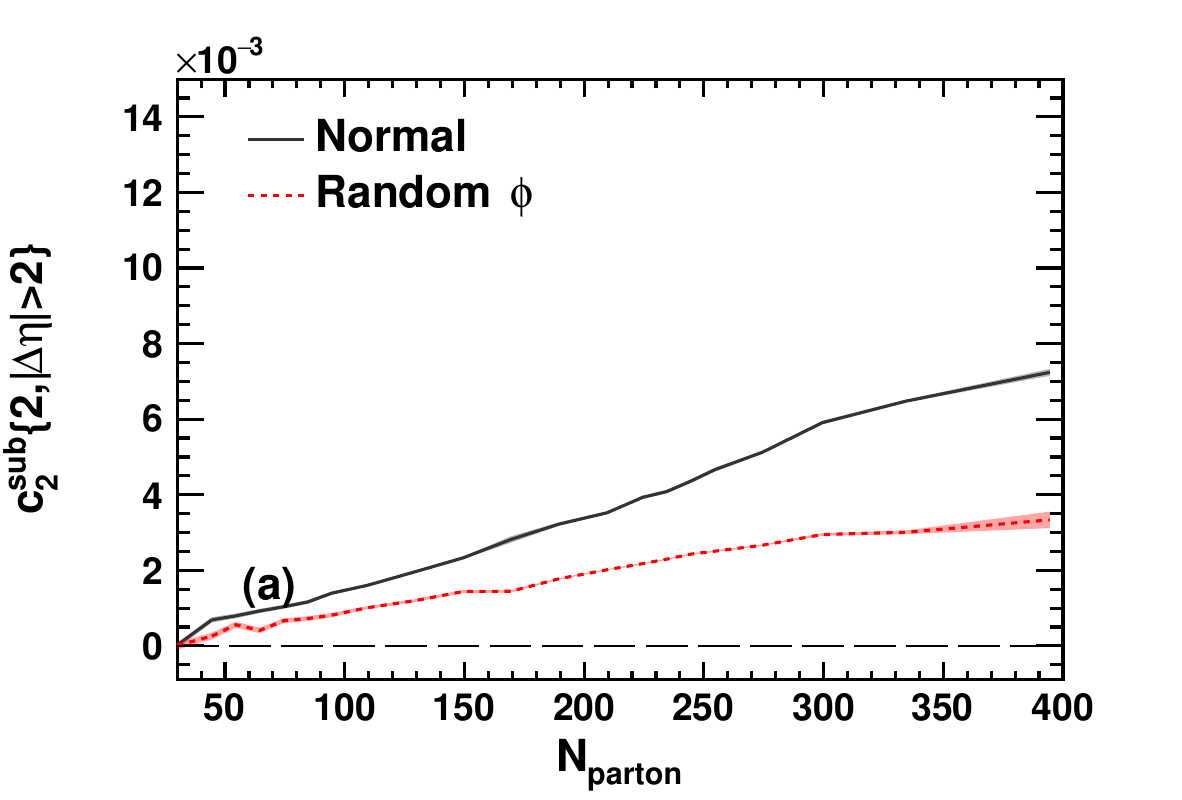}
	\includegraphics[width=0.49\textwidth]{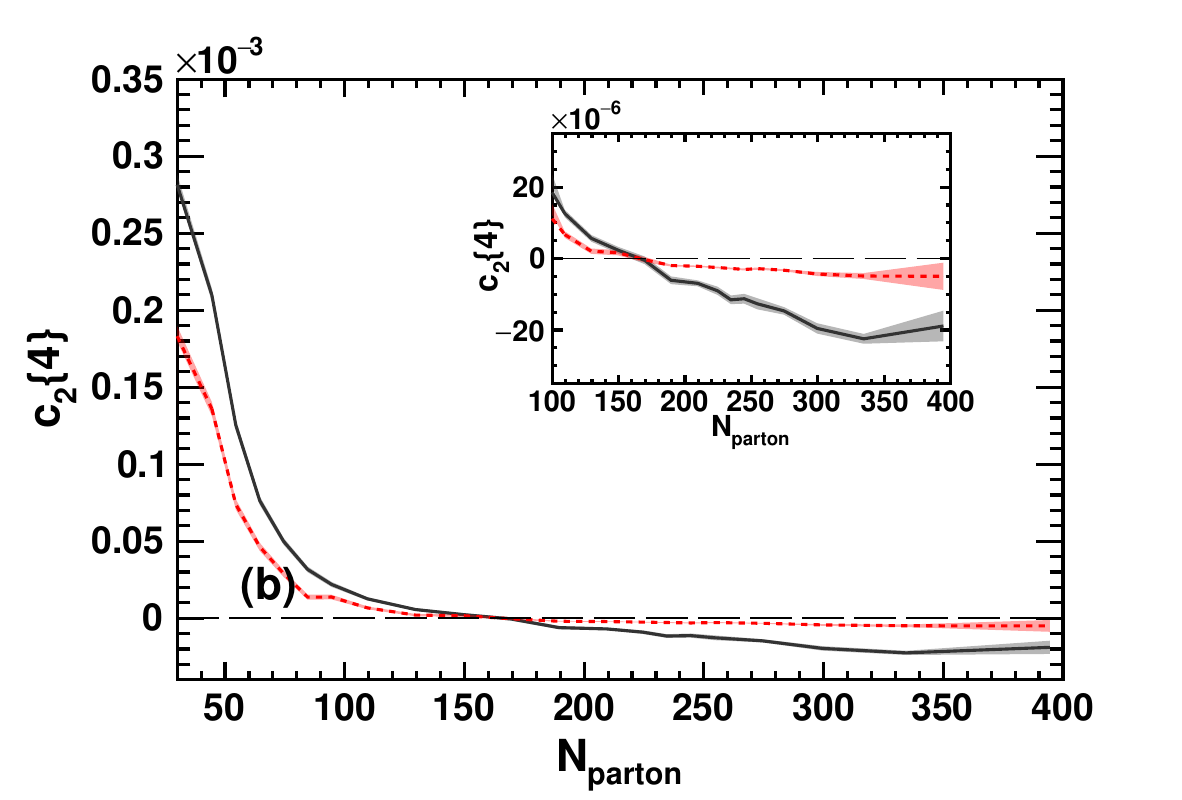}
	\caption{(Colour Online) Two-particle cumulants in the sub-event method after the low multiplicity events non-flow contribution is subtracted(a) and four-particle cumulants (b) for final state partons with $|\eta|<2.4$ and $0.3<p_{T}<3$ GeV$/c$ in pp collisions at $\sqrt{s}=13$ TeV versus the number of reference partons in the flow calculations. Gray solid lines represent the normal calculations while red dashed lines represent the results with random $\phi$ test. The bands represent the statistical uncertainties.}
	\label{fig:c2c4_randPhi}    
\end{figure*}

\begin{figure*}[hbt]
	\centering
	\includegraphics[width=0.49\textwidth]{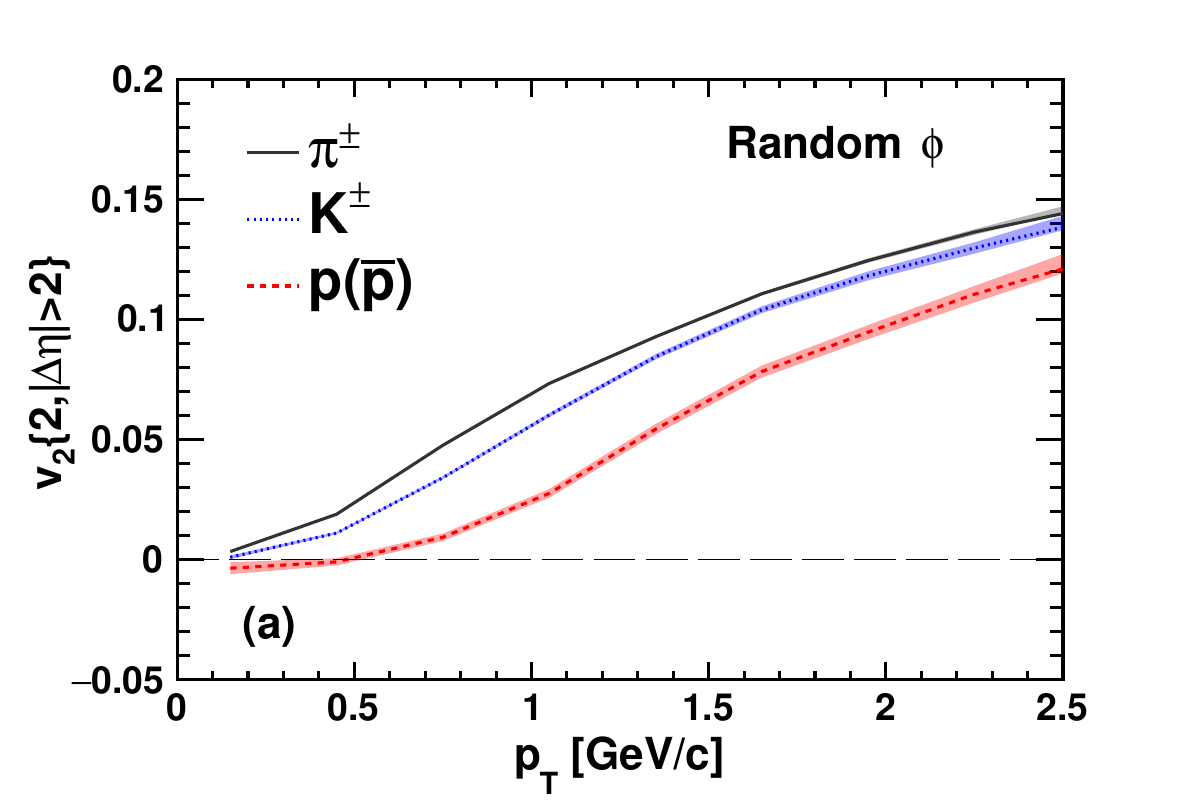}
	\includegraphics[width=0.49\textwidth]{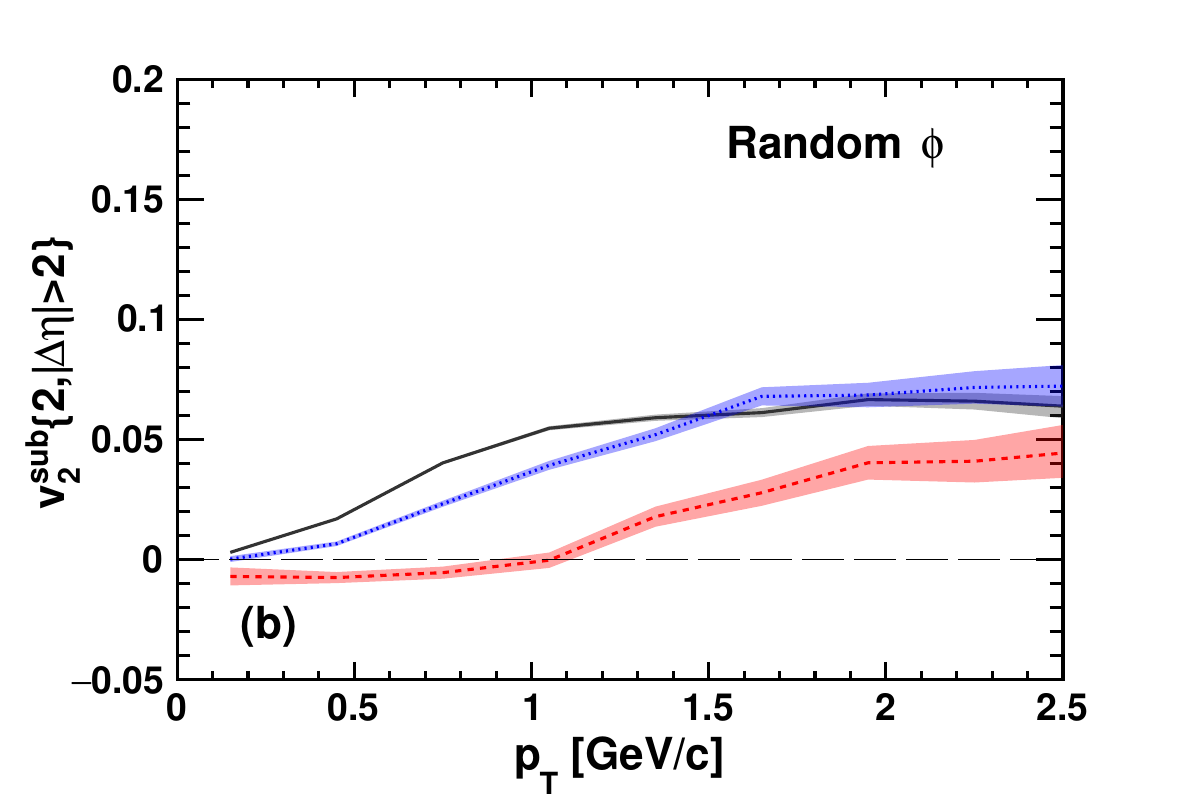}
	\caption{(Colour Online) The \pt differential \vtwo (a) and $\vtwo^{sub}$ (b) of identified hadrons with random $\phi$ test based on the sub-event method using pseudorapidity gap $|\deta|>2$ for high multiplicity events with $M>100$ in pp collisions at $\sqrt{s}=13$ TeV. The $\pi^{\pm}$, $K^{\pm}$, and proton results are shown by gray solid lines, blue dotted lines and red dashed lines, respectively. The bands represent the statistical uncertainties.}
	\label{fig:v2_piKp_randPhi}    
\end{figure*}

\subsection{Parton escape effects}
\label{sec:discussion}

As is known that the parton escape mechanism plays an important role in the flow development especially for small systems~\cite{He:2015hfa,Li:2016ubw,Li:2016flp}, we perform a random $\phi$ test calculation to understand the contribution of the escape mechanism in high energy pp collisions. In this test, the hydrodynamic type collective flow will be removed via randomizing the azimuthal angles of the partons after each parton-parton rescattering. We investigate the two-particle and four-particle cumulants at parton level in Fig.~\ref{fig:c2c4_randPhi}. The cumulants are studied with respect to $N_{parton}$ the number of partons having $0.3<p_{T}<3$ GeV$/c$ and $|\eta|<2.4$. We also perform a non-flow subtraction using the low multiplicity events with $20<M<40$ in Fig.~\ref{fig:c2c4_randPhi}(a).  It can be found in Fig.~\ref{fig:c2c4_randPhi}(a) that the amount of subtracted two particle cumulants which can be explained by the parton escape effects are generally sizable across the entire multiplicity region. With the increase of the multiplicity, the parton escape effect declines and accounts for only about 50\% of the total \ctwo cumulant.  The \cfour cumulant after randomization is entirely coming from parton escape. It is found in Fig.~\ref{fig:c2c4_randPhi}(b) that the escape mechanism induced \cfour is close to the trend of the normal \cfour and becomes negative around the same event multiplicity. This phenomenon suggests that parton escape can be important to understand the sign of the \cfour and offers us a new perspective to solve the sign puzzle of \cfour found in hydrodynamic studies~\cite{Zhao:2017rgg,Zhao:2020pty}.

In Fig.~\ref{fig:v2_piKp_randPhi}, we also examine the mass splitting of \vtwo for pion, kaon and proton with only parton escape effect when hadronic rescatterings are excluded. Clear mass hierarchy of low \pt flow is observed for the hadrons in the randomized $\phi$ test without and with the low multiplicity non-flow subtraction shown by Fig.~\ref{fig:v2_piKp_randPhi}(a) and Fig.~\ref{fig:v2_piKp_randPhi}(b), reinforcing the conclusion that the mass ordering of \vtwo is primarily developed during the coalescence process in high energy pp collisions within the transport model framework.

The AMPT model produces the particle species dependent flow in high energy proton proton collisions mostly during the coalescence process coupled to the parton evolution stage with some modifications during the hadron rescattering process. It is interesting to see that the mass ordering effect is also observed in hydrodynamic model implementations taking final state parton evolution stage into consideration~\cite{Zhao:2017rgg,Zhao:2020pty,YuanyuanWang:2023meu}, while it is only weakly generated in the color glass condensate model which relies on the initial state correlations~\cite{Schenke:2016lrs} or the pure hadronic interaction model~\cite{Zhou:2015iba}. This observation infers that the appearance of a significant mass ordering structure can be exclusively related to the creation of a deconfined parton matter phase in the high energy pp collisions~\cite{Zhang:2021ygs}. Fully examining this feature in the experimental data will shed a light on our understanding to the properties of the nuclear matter created in small collision systems.

\section{Summary}
\label{sec:summary}

Using a multi-phase transport model with PYTHIA8 initial conditions embedded with the sub-nucleon spatial fluctuations using three constituent quarks, we can successfully reproduce the charged hadron productions and the key features of the second order collective flow behavior observed in pp collisions at $\sqrt{s}=13$ TeV. By turning on the parton and hadron evolutions independently, we disentangle how the collective flow develops in different final state evolution stages of pp collisions. We find that the parton evolution stage is necessary for the creation of a sizable two-particle elliptic flow coefficient $v_{2}\{2\}$ and the negative four-particle cumulant $c_{2}\{4\}$. We also find that the coalescence process coupled to the final state parton evolution stage will predict a significant mass splitting feature of the \pt differential elliptic flow for identified hadrons. This strong mass ordering effect can be regarded as an important signature to reveal the existence of deconfined parton matter in high multiplicity pp collisions, considering that other approaches without final state parton evolutions usually deliver only small $v_2$ splittings. Therefore, investigating the mass ordering feature in experiments with high precision will be essential to distinguish whether this deconfined parton matter is formed in pp collisions~\cite{ALICE_talk}


 
We find that the collective flow in small systems also receives significant contributions from off-equilibrium effects like the parton escape mechanism. We perform a random $\phi$ test to isolate the escape effect in the development of collective flow in high energy pp collisions. We observe that the escape mechanism accounts for a large fraction of the two particle cumulant achieved in the final state. Interestingly, the appearance of a negative $c_{2}\{4\}$ in high multiplicity pp events is also found to be connected to the escape mechanism during the parton cascade. These observations demonstrate the necessity of considering the off-equilibrium effects when extracting the properties of the parton matter created in pp collisions.  
 
With the unique capability of including different mechanisms in a unified framework, the transport model approach presented in this study can be an important phenomenological tool to make quantitative comparisons to the collective flow  observables measured in small collision systems. This study may pave the way for further investigations on the azimuthal anisotropy measurements in small systems and help us to understand their origins.

\begin{acknowledgement}
We are grateful to Zhenyu Chen, Xinli Zhao, Shusu Shi, Siyu Tang, Chao Zhang, Xiaoming Zhang and Daimei Zhou for helpful discussions. This work is supported by the National Natural Science Foundation of China (Nos. 11905188, 12275103, 12061141008, 12322508, 12147101), the National Science Foundation under Grant No. 2012947 and 2310021 (Z.-W. L.), the Strategic Priority Research Program of Chinese Academy of Sciences (No. XDB34030000), the STCSM (No. 23590780100), the Fundamental Research Funds for the Central Universities China University of Geosciences (Wuhan) with No. G1323523064. 
\end{acknowledgement}

\bibliographystyle{epjstyle}   
\bibliography{main}   

\end{document}